\documentclass[sigconf]{acmart}
\usepackage{algorithm}
\usepackage{mathtools}
\usepackage{algorithmic}
\usepackage[font=small,skip=3pt]{caption}

\copyrightyear{2024}
\acmYear{2024}
\setcopyright{rightsretained}
\acmConference[IoT 2024]{14th International Conference on the Internet of Things}{November 19--22, 2024}{Oulu, Finland}
\acmBooktitle{14th International Conference on the Internet of Things (IoT 2024), November 19--22, 2024, Oulu, Finland}
\acmPrice{}
\acmDOI{10.1145/3703790.3703831}
\acmISBN{979-8-4007-1285-2/24/11}

\begin{document}

\title{Resource Slicing through Intelligent Orchestration of Energy-aware IoT services in Edge-Cloud Continuum}

\author{Hafiz Faheem Shahid}
\affiliation{%
  \institution{University of Oulu}
  \city{Oulu}
  \country{Finland}}
\email{faheem.shahid@oulu.fi}
\orcid{0009-0005-3350-0276}

\author{Erkki Harjula}
\affiliation{%
  \institution{University of Oulu}
  \city{Oulu}
  \country{Finland}}
\email{erkki.harjula@oulu.fi}
\orcid{0000-0001-5331-209X}

\begin{abstract}

The rapid growth of the Internet of Things (IoT) applications inflicts high requirements for computing resources and network bandwidth. A growing number of service providers are applying edge-cloud computing to improve the quality of their services. Deploying IoT applications to optimal computing nodes to minimize energy consumption and enhance system performance remains an open challenge. In this paper, we present an intelligent orchestration concept for breaking down IoT applications into granular microservices, called nanoservices, and deploying them in an energy-aware manner to optimal computing nodes in the edge-cloud continuum by applying resource and network slicing methods. With this consolidated slicing scheme, we can efficiently allocate network and compute resources to meet the needs of these nanoservices.


\end{abstract}


\maketitle

\vspace{-2mm}
\section{Introduction}
Network slicing is an essential 5G solution to meet individual requirements and performance needs of various applications and services. A network slice can be considered as a collection of network flows that belong to multiple devices with different resource capacity on the edge-cloud continuum and users \cite{baktir2024end}. The concept of semantic slicing extends the idea to generic semantic optimization of resources, data processing and task scheduling in the edge-cloud computing continuum \cite{10287519}. It leverages deep data understanding and application-specific requirements to intelligently allocate resources across the computing continuum, creating efficient, adaptive systems for diverse applications. Resource slices in semantic slicing can take various forms, addressing computational data from specific devices like sensors or actuators, IoT applications, or end users\cite{richart2016resource}. In the world of depleting natural resources, resource efficiency — particularly energy efficiency — is becoming as important as Quality of service (QoS) in the design of all online services. This paper addresses the open challenge of deploying services in the computing continuum in an energy-aware manner, while meeting required QoS through resource-aware orchestration of semantic slicing.





\vspace{-2mm}
\section{Background \& Related work}
Virtualization has enabled building flexible microservice-based architectures that can operate across geographical and logical locations, allowing distributing service components - microservices - dynamically over diverse environments  \cite{harjula2019decentralized}. This flexibility supports rapid scaling and adaptation, i.e. microservices can be deployed and managed independently across cloud, edge, and local nodes, optimizing for performance, latency, and resource availability at each location. Building on the basic concept of microservice architectures, our nanoservice approach \cite{harjula2019decentralized} makes it possible to break IoT services into granular, lightweight parts, called "nanoservices," which deployed on local constrained-capacity nodes, as well as on more powerful edge and cloud nodes. This setup provides a flexible service architecture that spans local, edge, and cloud tiers, enabling tasks to be deployed on the optimal nodes. Combined with network slicing, this concept forms the foundation for "semantic slicing", i.e. customized service layers that adapt to application and service-specific needs. Adding resource-awareness \cite{taleb2017multi}, these services can be orchestrated in ways that save energy and optimize resource use, making IoT services efficient to meet QoS requirements.

\vspace{-2mm}
\section{Methods \& Technical Approaches}
We propose an Intelligent orchestration system that splits an IoT application into a set of multiple nanoservices and distributes this operation into the computing continuum.
Network slicing and resource slicing are the two main components to perform the optimization of resource and energy-efficiency by utilizing energy forecasting and resource-aware deployment functionalities. When combining with resource and QoS-aware orchestration, we can enable deploying IoT applications and service components in an energy-aware manner, while ensurig their sufficient performance.

\vspace{-2mm}
\subsection{Resource Slicing}
\textbf{Nanoservice Decomposition:} The concept of our proposed architecture divides IoT applications into multiple nanoservices, where each nanoservice performs different operations and tasks like sensing the environment, performing heavy data processing and providing quick responses for time critical tasks.

\begin{figure}[htbp]
  \vspace{-1.5mm}
  \centering
  \includegraphics[trim=0.1in 0.1in 0.05in 0.05in, clip, width=0.95\linewidth]{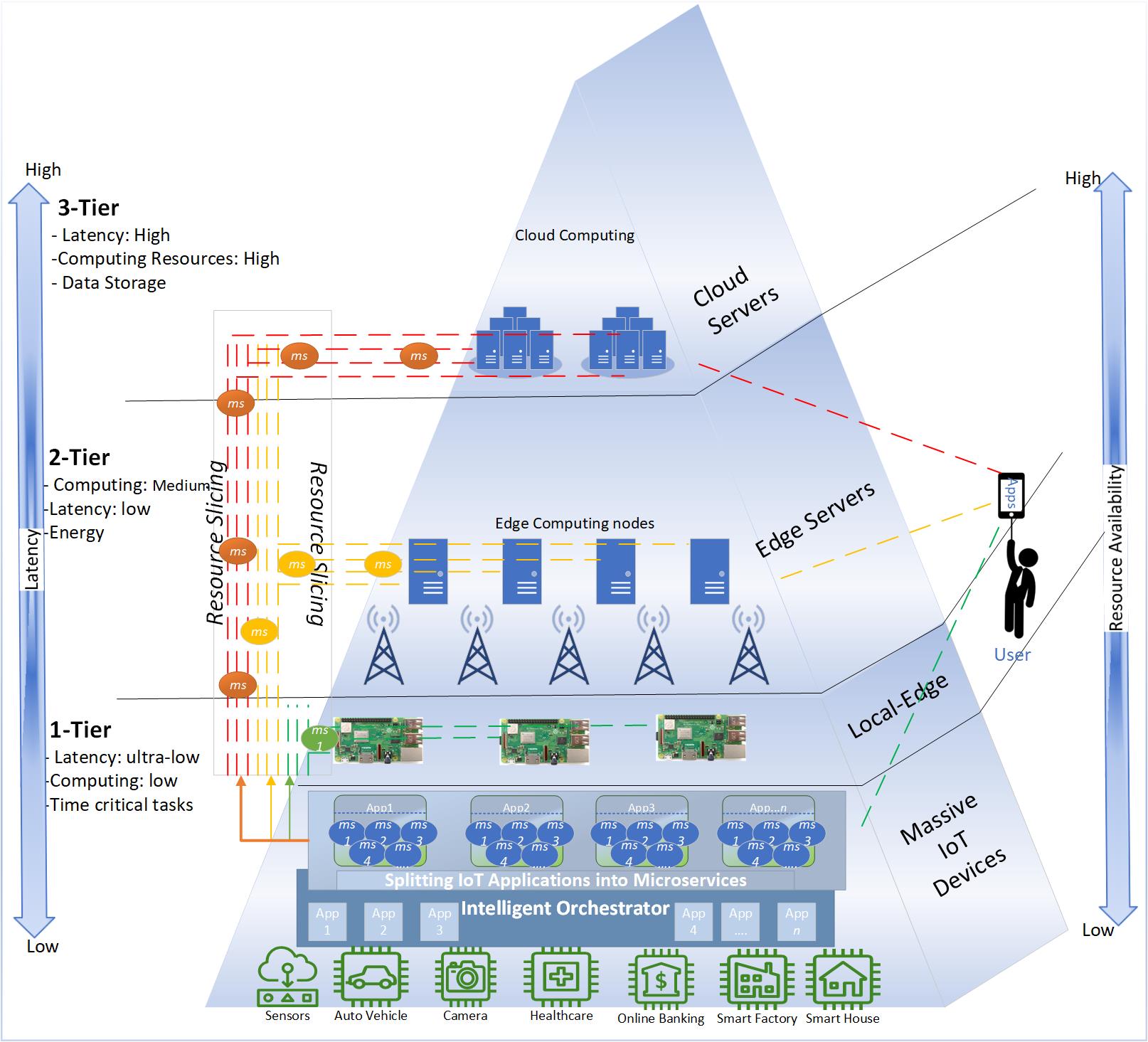}
  \captionsetup{font=small}
  \caption{Semantic slicing in multi-tier computing continuum.}
  \label{Resource Slicing in Multi-Tier Computing Architecture}
  \vspace{-3.5mm}
\end{figure}

\noindent
\textbf{Resource Allocation:} Each nanoservice has 
a
set of resource requirements concerning e.g. the needed Central Processing Unit (CPU), Graphics Processing Unit (GPU), memory or storage capacity. Furthermore, the IoT applications have performance and efficiency requirements related to e.g. network latency, computational performance or energy consumption. The proposed architecture aims to deploy each microservice in a manner making sure that these requirements are met. As an example, we can reduce the inflicted energy cost by utilizing energy forecast to decide on which devices and at which time a task should be performed. Fig.\ref{Resource Slicing in Multi-Tier Computing Architecture} illustrates the deployment of these tasks encapsulated into nanoservices across the three tiers of the computing continuum. A cloud node, for example, can manage computationally heavy tasks with no real-time requirements (red dashed line), while an edge node with maximum available GPUs can host tasks requiring intensive data exchange and low latency with the end-node (yellow dashed line). Local node with sufficient computational capacity
can host lightweight tasks requiring low latency with the end-node (green dashed line).
\vspace{-2.08mm}

\vspace{-1.5mm}
\subsection{Network Slicing}
\textbf{Network Requirements of Nanoservices:} Nanoservices may have varying communication requirements depending on the type of operations they are performing. For instance, a task that performs real-time monitoring, may require high bandwidth and low latency, whereas a task for processing and managing big data would require high bandwidths but would be more delay-tolerant.

\noindent\textbf{Network Slicing Approach:} In our approach, we use the concept of network slicing technique to split the network infrastructure into various virtual networks (VN).
For instance, an energy-intensive nanoservice that manages real-time data and is extremely dependent on the latency, could be deployed to Ultra-Reliable Low Latency Communication (URLLC) slice. Conversely, a nanoservice requiring high network throughput, would be deployed to Enhanced Mobile Broadband (eMBB) slice, while delay-tolerant nanoservice dealing with low data rates could be deployed to massive Machine Type Communications (mMTC) slice.

\vspace{-2mm}
\subsection{Energy-aware Orchestration}

\begin{figure}[htbp]
  \centering
  \includegraphics[width=\linewidth]{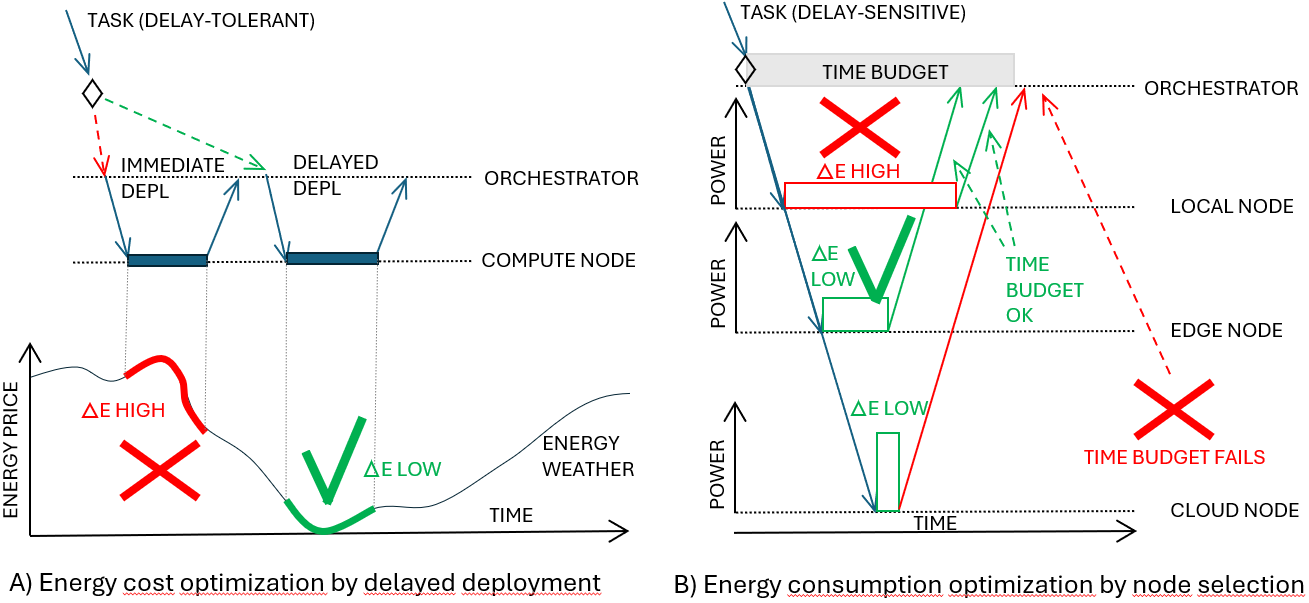}
  \captionsetup{font=small}
  \caption{Energy optimization by energy-aware orchestration.}
  \label{Example_Orchestration}
  \vspace{-3mm}
\end{figure}

In the hypothetical scenario of Fig.\ref{Example_Orchestration}, we illustrate the intelligent orchestration concept in action. Fig.\ref{Example_Orchestration}A, illustrates energy cost optimization based on energy price forecast by delaying the deployment of a delay-tolerant task to a time when the energy cost is estimated to be lower. Fig.\ref{Example_Orchestration}B, for one illustrates energy consumption optimization by continuously selecting the most energy-efficient node based on their energy consumption profiles, continuously updated by, e.g., reinforcement learning. Here, the orchestrator calculates that the local computing node has the lowest power consumption, but its long processing time leads to the highest accumulated energy usage. The cloud deployment, for one, is estimated to have low accumulated energy consumption, but it fails to meet the task's timing constraints. Therefore, the edge deployment, giving the lowest cumulative energy consumption, and meeting the timing requirements, is selected for the deployment.

\vspace{-2mm}
\section*{Conclusion}
In this paper, we introduced an intelligent orchestration concept for energy-aware IoT applications, which uses semantic slicing, including resource and network slicing methods to optimally deploy computing tasks on the computing nodes. We simplified resource allocation and execution of IoT application in edge computing environments by handling each microservice as a single task with diverse set of resource and communication requirements. While this short paper focused on introducing the basic concept, the future work includes evaluating and validating the performance of the concept with AI algorithms in both simulated and real-world setups with 6G Flagship and Eware-6G projects\footnote{\textbf{Acknowledgments:} This  work is supported by Eware-6G project funded by Business Finland (grant 8819/31/2022), and 6G Flagship program funded by the Research Council of Finland (grant 346208).}. 


\bibliographystyle{ACM-Reference-Format}
\bibliography{sample-base}

\end{document}